\documentstyle[12pt]{article}
\begin{document}

\hfill DUKE-CGTP-2000-15

\hfill hep-th/0008184

\vspace*{2.0in}

\begin{center}

{\large\bf Discrete Torsion in }

{\large\bf Perturbative Heterotic String Theory}

\vspace{1in}

Eric Sharpe \\
Department of Physics \\
Box 90305 \\
Duke University \\
Durham, NC  27708 \\
{\tt ersharpe@cgtp.duke.edu} \\

$\,$ \\

\end{center}

In this paper we analyze discrete torsion in perturbative
heterotic string theory.  In previous work we have given a purely
mathematical explanation of discrete torsion as the choice of
orbifold group action on a $B$ field, in the case that $d H = 0$; in this
paper we perform the analogous calculations in heterotic strings
where $d H \neq 0$.

\begin{flushleft}
August 2000 
\end{flushleft}

\newpage

\tableofcontents

\newpage

\section{Introduction}

Discrete torsion is a historically-mysterious degree of freedom
associated with orbifolds, originally discovered in \cite{vafa1}.
In previous work, we explained discrete torsion 
for type II $B$ fields \cite{dt1,dt2,dt3} (as well
as the M-theory three-form potential $C$ \cite{cdt}).
To summarize our results, we found that
\begin{center}
{\it Discrete torsion is the choice of orbifold group action on the $B$
field.}
\end{center}
In particular, we showed that discrete torsion has nothing
to do with string theory {\it per se}, but rather has a purely
mathematical understanding.

However, in our previous work \cite{dt1,dt2,dt3} we assumed
that the curvature of the $B$ field, namely $H$, satisfied the usual
Bianchi identity $d H = 0$.  Unfortunately this is not the case
for heterotic $B$ fields, where (as is well-known) $d H = 
\mbox{Tr } F \wedge F - \mbox{Tr } R \wedge R$.
So, strictly speaking, the results of \cite{dt1,dt2,dt3} do not apply
to the case of the heterotic $B$ field.

In this short paper we shall fill this gap in our understanding
by examining orbifold group actions on heterotic $B$ fields.
At the end of the day, we find that the difference between any two
orbifold group actions on a heterotic $B$ field is defined by
the same data as in \cite{dt1,dt2,dt3} -- so although heterotic
$B$ fields look somewhat different from type II $B$ fields,
and although orbifold group actions on heterotic $B$ fields are
twisted by comparison, the difference between any two orbifold
group actions can be described the same way for heterotic $B$ fields
as for type II $B$ fields.

We begin by working out a complete description of heterotic
$B$-fields on local coordinate charts.  Before we can accomplish
that goal, however, we first review relevant facts concerning
Chern-Simons forms in
section~\ref{chsrev}.  Once we understand Chern-Simons terms at
a sufficiently deep level, we work out a local-coordinate chart
description of heterotic $B$-fields in section~\ref{hetB}.

Once we understand heterotic $B$-fields sufficiently well,
we proceed to study orbifold group actions.  As heterotic $B$ fields
are tied to gauge and tangent bundles, we first study
orbifold group actions on principal $G$-bundles
with connection (for general $G$) in section~\ref{orbbnd}.
(In previous work \cite{dt1,dt2,dt3} we have exhaustively
discussed principal $G$-bundles with connection for 
$G$ abelian; here we describe the general case.)
Then, we discuss the induced orbifold group actions on 
Chern-Simons forms in section~\ref{orbgrb}.
Once we have the basics down, we use the usual self-consistent
bootstrap to work out orbifold group actions on
heterotic $B$-fields in section~\ref{orbhetB}.

Finally, in section~\ref{diffs} we conclude by discussing the differences
between orbifold group actions on heterotic $B$-fields.
We find that the differences between orbifold group actions
on heterotic $B$ fields (for fixed action on the gauge and tangent bundles)
is defined by the same data as for type II $B$ fields
\cite{dt1,dt2,dt3}, and so we recover the usual
$H^2(\Gamma, U(1))$, twisted sector phases of \cite{vafa1},
and so forth.  

This paper is a continuation of 
the papers \cite{dt3} and \cite{cdt}, and so readers are
encouraged to read them first.

\section{Review of Chern-Simons forms}    \label{chsrev}

Before we describe the heterotic $B$-field in local coordinate
patches, we shall first take a moment to review Chern-Simons forms.

For simplicity, we shall assume that $\mbox{Tr } F \wedge F$ is normalized
to be (the image of) an integral cohomology class.
Assume that $F$ is a connection on a principal $G$-bundle
with transition functions $g_{\alpha \beta}$ (defined with respect
to some good cover), and let $A^{\alpha}$ denote the 
connection (the gauge field) in patch $U_{\alpha}$.
On overlaps,
\begin{displaymath}
A^{\alpha} \: = \: g_{\alpha \beta} \, A^{\beta} \, g^{-1}_{\alpha \beta}
\: - \: \left( d g_{\alpha \beta} \right) g^{-1}_{\alpha \beta} 
\end{displaymath}
To set conventions, define $F = d A + A \wedge A$,
then it is trivial to verify that $F^{\alpha} = 
g_{\alpha \beta} F^{\beta} g_{\alpha \beta}^{-1}$,
and so $\mbox{Tr } F^{\alpha} \wedge F^{\alpha} = 
\mbox{Tr } F^{\beta} \wedge F^{\beta}$.

Now, given some form that lies in the image of integral cohomology,
in principle one can construct the other elements of a \v{C}ech-de
Rham cocycle.  The first step in this is well-known:
\begin{displaymath}
\mbox{Tr } F^{\alpha} \wedge F^{\alpha} \: = \:
d \, \omega_3^{\alpha}
\end{displaymath}
where
\begin{displaymath}
\omega_3^{\alpha} \: = \: \mbox{Tr }\left( A^{\alpha} \wedge d A^{\alpha} 
 \: + \: 
\frac{2}{3} A^{\alpha} \wedge A^{\alpha} \wedge A^{\alpha} \, \right)
\end{displaymath}
is the usual Chern-Simons three-form.

The second step is a little more obscure, but can also be worked out.
Note that
\begin{displaymath}
\omega_3^{\alpha} \: - \: \omega_3^{\beta} \: = \:
- \, \mbox{Tr } \left( g_{\alpha \beta}^{-1} ( d g_{\alpha \beta} ) \wedge
d A^{\beta} \right)
\end{displaymath}
Since $g_{\alpha \beta}^{-1} d g_{\alpha \beta}$ is a closed form,
and we are working on a good cover, there exists a function 
$\Lambda_{\alpha \beta}$ such that
$g_{\alpha \beta}^{-1} d g_{\alpha \beta} = d \Lambda_{\alpha \beta}$,
and so we can write
\begin{displaymath}
\omega_3^{\alpha} \: - \: \omega_3^{\beta} \: = \: d \omega_2^{\alpha \beta}
\end{displaymath}
where
\begin{displaymath}
\omega_2^{\alpha \beta} \: = \: - \mbox{Tr } \left(
\Lambda_{\alpha \beta} d A^{\beta} \right)
\end{displaymath}

In addition, there also exist 
local 1-forms $\omega_1^{\alpha \beta \gamma}$ and local functions
$h_{\alpha \beta \gamma \delta}$ filling out the rest of the
\v{C}ech-de Rham cocycle.  We can summarize this data as follows:
\begin{eqnarray*}
\mbox{Tr } F^{\alpha} \wedge F^{\alpha} & = & d \omega_3^{\alpha} \\
\omega_3^{\alpha} \: - \: \omega_3^{\beta} & = &
d \omega_2^{\alpha \beta} \\
\omega_2^{\alpha \beta} \: + \: \omega_2^{\beta \gamma} \: + \:
\omega_2^{\gamma \alpha} & = & d \omega_1^{\alpha \beta \gamma} \\
\omega_1^{\beta \gamma \delta} \: - \:
\omega_1^{\alpha \gamma \delta} \: + \:
\omega_1^{\alpha \beta \delta} \: - \:
\omega_1^{\alpha \beta \gamma} & = & d \log h_{\alpha \beta \gamma
\delta} \\
\delta h_{\alpha \beta \gamma \delta} & = & 1
\end{eqnarray*}

Somewhat more formally, we have described $\mbox{Tr }F \wedge F$
as the curvature of a 2-gerbe associated to the principal $G$-bundle
with connection.

This discussion is somewhat complicated, but a simpler version
also exists for $\mbox{Tr }F$.
We can write
\begin{eqnarray*}
\mbox{Tr } F^{\alpha} & = & d \mbox{Tr } A^{\alpha} \\
\mbox{Tr } A^{\alpha} \: - \: \mbox{Tr } A^{\beta} & = &
- \mbox{Tr } \left( (d g_{\alpha \beta}) g^{-1}_{\alpha \beta} \right) \\
 & = & d \left( \mbox{Tr } \log g_{\alpha \beta} \right) \\
 & = & d \left( \log \det g_{\alpha \beta} \right) \\
\delta \left( \det g_{\alpha \beta} \right) & = & 1
\end{eqnarray*}
Formally, we have described $\mbox{Tr }F$ as the curvature of a 0-gerbe
(a principal $U(1)$ bundle with connection) associated to the 
principal $G$-bundle with connection.
In fact, this associated 0-gerbe is precisely the determinant bundle.

\section{Heterotic $B$-fields}  \label{hetB}

We are now ready to discuss the $B$ field in perturbative heterotic
strings.  First recall that the curvature $H$ of the $B$ field obeys
\begin{displaymath}
d H \: = \: \mbox{Tr } F \wedge F \: - \: \mbox{Tr } R \wedge R
\end{displaymath}
With this in mind, to each open set $U_{\alpha}$ in a good cover,
we associate a three-form $H^{\alpha}$ and a two-form $B^{\alpha}$
related as
\begin{displaymath}
H^{\alpha} \: = \: d B^{\alpha} \: + \:
\omega_{3, F}^{\alpha} \: - \: \omega_{3, R}^{\alpha} 
\end{displaymath}

Next, how are the $B$ fields on overlapping patches related?
Recall that as part of the Green-Schwarz anomaly cancellation mechanism,
gauge transformations of either the gauge bundle or the tangent bundle
induce gauge transformations of $B$.  Specifically,
if under a gauge transformation
\begin{displaymath}
\omega_{3, F}^{\alpha} \: \mapsto \: \omega_{3, F}^{\alpha} 
\: - \: \mbox{Tr } \left(
d \Lambda \wedge d A^{\alpha} \right)
\end{displaymath}
then one must simultaneously have
\begin{displaymath}
B^{\alpha} \: \mapsto \: B^{\alpha} \: + \: \mbox{Tr } \left(
\Lambda d A^{\alpha} \right)
\end{displaymath}
so that $H^{\alpha}$ remains invariant.
Since the connections on the gauge and tangent bundles on
overlapping patches are related by gauge transformations
(defined by the transition functions), we find that in general,
the difference between two-forms $B^{\alpha}$ on overlapping
patches is given by
\begin{equation}    \label{boverlap}
B^{\alpha} \: - \: B^{\beta} \: = \:
d A^{\alpha \beta} \: - \: \omega_{2, F}^{\alpha \beta} \: + \:
\omega_{2, R}^{\alpha \beta} 
\end{equation}
for some local one-forms $A^{\alpha \beta}$.

Note that as a consequence of the expression above,
$H^{\alpha} = H^{\beta}$ on overlapping patches, i.e.,
$H^{\alpha} = H |_{U_{\alpha}}$ for some globally-defined three-form
$H$.

Next, adding the expressions~(\ref{boverlap}) on each double overlap
in a triple overlap, we are forced to conclude that
\begin{equation}   \label{atriple}
A^{\alpha \beta} \: + \:
A^{\beta \gamma} \: + \: A^{\gamma \alpha} \: = \:
\omega_{1, F}^{\alpha \beta \gamma} \: - \:
\omega_{1, R}^{\alpha \beta \gamma} \: + \:
d \log h^B_{\alpha \beta \gamma}
\end{equation}
for some $U(1)$-valued
functions $h^B_{\alpha \beta \gamma}$ defined on triple overlaps.

Finally, from adding the expressions~(\ref{atriple}) on each
triple overlap in a quadruple overlap, we are forced to conclude that
\begin{equation}
\left( h^B_{\beta \gamma \delta} \right) \,
\left( h^B_{\alpha \gamma \delta} \right)^{-1} \,
\left( h^B_{\alpha \beta \delta} \right) \,
\left( h^B_{\alpha \beta \gamma} \right)^{-1}
\: = \:
\left( h^F_{\alpha \beta \gamma \delta} \right)^{-1} \,
\left( h^R_{\alpha \beta \gamma \delta} \right)
\end{equation}
Note this means that at the level of \v{C}ech cohomology,
the 3-cochains $h^F_{\alpha \beta \gamma \delta}$ and
$h^R_{\alpha \beta \gamma \delta}$ are cohomologous;
their difference is a coboundary defined by the $h^B_{\alpha \beta \gamma}$.

To summarize, we have found that the heterotic $B$ field
is described, in local coordinate patches, 
by a globally-defined three-form $H$, local two-forms
$B^{\alpha}$, local one-forms $A^{\alpha \beta}$,
and local $U(1)$-valued functions $h^B_{\alpha \beta \gamma}$
obeying
\begin{eqnarray*}
d H & = & \mbox{Tr } F \wedge F \: - \:
\mbox{Tr } R \wedge R \\
H |_{U_{\alpha}} & = & d B^{\alpha} \: + \: \omega_{3, F}^{\alpha}
\: - \: \omega_{3, R}^{\alpha} \\
B^{\alpha} \: - \: B^{\beta} & = &
d A^{\alpha \beta} \: - \: \omega_{2, F}^{\alpha \beta} \: + \:
\omega_{2, R}^{\alpha \beta} \\
A^{\alpha \beta} \: + \: A^{\beta \gamma} \: + \: 
A^{\gamma \alpha} & = &
\omega_{1, F}^{\alpha \beta \gamma} \: - \:
\omega_{1, R}^{\alpha \beta \gamma} \: + \:
d \log h^B_{\alpha \beta \gamma} \\
\left( h^B_{\beta \gamma \delta} \right) \,
\left( h^B_{\alpha \gamma \delta} \right)^{-1} \,
\left( h^B_{\alpha \beta \delta} \right) \,
\left( h^B_{\alpha \beta \gamma} \right)^{-1}
& = & 
\left( h^F_{\alpha \beta \gamma \delta} \right)^{-1} \,
\left( h^R_{\alpha \beta \gamma \delta} \right) 
\end{eqnarray*} 
More formally, the heterotic $B$ field defines a map
between the 2-gerbes with connection associated to the gauge and
tangent bundles.

\section{Orbifold group action on principal $G$-bundles}   \label{orbbnd}

In prior work \cite{dt1,dt3} we have exhaustively discussed
orbifold group actions on principal $U(1)$ bundles.
In order to discuss orbifold group actions in heterotic
string theory, however, we need to examine orbifold group actions
on principal $G$-bundles for more general Lie groups $G$.

To set conventions, assume we have a bundle with connection
described by Ad($G$)-valued gauge fields $A^{\alpha}$ 
(one for each element $U_{\alpha}$ of a ``good invariant'' cover,
as described in \cite{dt1,dt3})
and transition functions $g_{\alpha \beta}$, obeying
\begin{eqnarray*}
A^{\alpha} & = & g_{\alpha \beta} \, A^{\beta} \, g^{-1}_{\alpha \beta}
\: - \: \left( d g_{\alpha \beta} \right) g^{-1}_{\alpha \beta} \\
g_{\alpha \beta} \, g_{\beta \gamma} \, g_{\gamma \alpha} & = & 1
\end{eqnarray*}

Proceeding as in \cite{dt1,dt2,dt3}, define \v{C}ech cochains
$\gamma^g_{\alpha}$ by
\begin{equation}
g^* g_{\alpha \beta} \: = \:
\left( \gamma^g_{\alpha} \right) \,
\left( g_{\alpha \beta} \right) \,
\left( \gamma^g_{\beta} \right)^{-1}
\end{equation}
From expanding $(g_1 g_2)^* g_{\alpha \beta}$ in two different ways,
we are led to demand
\begin{equation}     \label{gammaconstr}
\gamma^{g_1 g_2}_{\alpha} \: = \:
\left( g_2^* \gamma^{g_1}_{\alpha} \right) \,
\left( \gamma^{g_2}_{\alpha} \right)
\end{equation}
and from demanding consistency of $A^{\alpha}$ on overlaps,
we are led to derive (as in \cite{dt3})
\begin{equation}
g^* A^{\alpha} \: = \:
\left( \gamma^g_{\alpha} \right) \, A^{\alpha} \, 
\left( \gamma^g_{\alpha} \right)^{-1} \: + \:
\left( \gamma^g_{\alpha} \right) \,
d \left( \gamma^g_{\alpha} \right)^{-1}
\end{equation}

Now, in \cite{dt1,dt2,dt3} we pointed out that both orbifold $U(1)$
Wilson lines and discrete torsion arise as the differences between
orbifold group actions.  (Put another way, the set of orbifold group
actions is only a set in general, not a group, but it is acted upon by
a group in those cases.)
Let us attempt to repeat that analysis here.
Let $\left( \gamma^g_{\alpha} \right)$,
$\left( \overline{\gamma}^g_{\alpha} \right)$ define a pair of
orbifold group actions on some principal $G$-bundle with connection,
as above.  Define
\begin{equation}
\phi^g_{\alpha} \: = \: \left( \overline{\gamma}^g_{\alpha} \right)^{-1} \,
\left( \gamma^g_{\alpha} \right)
\end{equation}
By expressing $g^* g_{\alpha \beta}$ in terms of these two actions, we find
\begin{equation}     \label{phiconstr}
\phi^g_{\alpha} \, g_{\alpha \beta} \: = \: g_{\alpha \beta} \,
\phi^g_{\beta}
\end{equation}
The expression above for $\left( \phi^g_{\alpha} \right)$
shows that $\left( \phi^g_{\alpha} \right)$ defines a base-preserving 
automorphism
of the principal $G$-bundle \cite[section 5.5]{husemoller}.
Base-preserving automorphisms of a principal $G$-bundle are 
gauge transformations, so this means that $\left( \phi^g_{\alpha} \right)$
defines a gauge transformation of the bundle.

The reader will probably be slightly confused to hear
that equation~(\ref{phiconstr}) implies that 
$\left( \phi^g_{\alpha} \right)$ defines a gauge transformation.
After all, one usually thinks of gauge transformations of bundles
as being global maps into $G$, and if $\left( \phi^g_{\alpha} \right)$
defines a global map, then one would expect that
$\phi^g_{\alpha} = \phi^g_{\beta}$ on $U_{\alpha} \cap U_{\beta}$,
not equation~(\ref{phiconstr}).
Unfortunately, working at the level of \v{C}ech cochains means
implicitly working in local trivializations, and for general $G$,
including local trivializations makes the relationship between
bundle automorphisms and gauge transformations less transparent.

So far we have argued that the difference between any two orbifold
group actions on a principal $G$ bundle is defined by a set of
gauge transformations.  This is very reminiscent of \cite{dt1,dt2,dt3}
where we argued that the difference between any two orbifold group
actions on a principal $U(1)$ bundle or on a $B$ field is defined
by a set of gauge transformations.  However, there is an important
difference in the present case -- although the difference between
any two orbifold group actions is a set of gauge transformations,
the gauge transformations do not form a representation of the
orbifold group in general.

Specifically, from equation~(\ref{gammaconstr}) we find that
\begin{equation}
\phi^{g_1 g_2}_{\alpha} \: = \:
\left( \overline{\gamma}^{g_2}_{\alpha} \right)^{-1} \,
\left( g_2^* \phi^{g_1}_{\alpha} \right) \,
\left( \gamma^{g_2}_{\alpha} \right)
\end{equation}
In order for the gauge transformations $\phi^g_{\alpha}$ to define
a representation of the orbifold group, we would have needed
$\phi^{g_1 g_2}_{\alpha} = \left( g_2^* \phi^{g_1}_{\alpha} \right)
\left( \phi^{g_2}_{\alpha} \right)$, but we see that this will
only be true if $\overline{\gamma}^{g_2}_{\alpha}$ commutes with
$g_2^* \phi^{g_1}_{\alpha}$, which will not be true in general.

However, in very special cases one can sometimes still
recover a description of orbifold Wilson lines for
principal $G$-bundles with connection in terms of
$\mbox{Hom}(\Gamma, G)/G$, the description most familiar
to physicists.  Specialize to the canonically
trivial bundle (i.e., $g_{\alpha \beta} \equiv 1$ for all $\alpha$, $\beta$)
over some path-connected space, with
connection identically zero.  On this principal $G$-bundle
with connection there is a canonical trivial orbifold group action, 
defined by taking $\gamma^g_{\alpha} \equiv 1$ for all $g \in \Gamma$
and all $\alpha$.  There is also a family of nontrivial orbifold
group actions, defined by taking $\gamma^g_{\alpha}$ to be constant
maps into $G$ (i.e., $\gamma^g_{\alpha} = \gamma^g |_{U_{\alpha}}$),
forming a representation of the orbifold group:
\begin{displaymath}
\gamma^{g_1 g_2} \: = \: \left( \gamma^{g_1}  \right) \,
\left( \gamma^{g_2} \right)
\end{displaymath}
In other words, each set of $\{ \gamma^g \}$ defining an orbifold
group action defines an element of $\mbox{Hom}(\Gamma, G)$.
The reader can easily check that such $\gamma^g$ yield a well-defined
orbifold group action on the canonically trivial principal $G$-bundle
with zero connection.

Now, we should be slightly careful.  Not all of the elements
of $\mbox{Hom}(\Gamma, G)$ define distinct orbifold group actions
on this special bundle with connection.  Under a constant gauge
transformation $\phi$, the connection transforms as
$A^{\alpha} \mapsto \phi A^{\alpha} \phi^{-1}$.
As a result, given an orbifold group action defined by
constant $\gamma^g$ as
\begin{displaymath}
g^* A^{\alpha} \: = \: \left( \gamma^g \right) \,
\left( A^{\alpha} \right) \,
\left( \gamma^g \right)^{-1}
\end{displaymath}
if we gauge-transform by constant $\phi$ we get
\begin{displaymath}
\phi \left( g^* A^{\alpha} \right) \phi^{-1} \: = \:
\left( \gamma^g \right) \,
\left( \phi A^{\alpha} \phi^{-1} \right) \,
\left( \gamma^g \right)^{-1}
\end{displaymath}
which can be rewritten as
\begin{displaymath}
g^* A^{\alpha} \: = \:
\left( \phi^{-1} \gamma^g \phi \right) \,
\left( A^{\alpha} \right) \,
\left( \phi^{-1} \gamma^g \phi \right)^{-1}
\end{displaymath}
In other words, a constant gauge transformation (on this special
bundle with connection) will map an orbifold group action
defined by $\left\{ \gamma^g \right\}$ to an orbifold group action defined
by $\left\{ \phi^{-1} \gamma^g \phi \right\}$.
Conversely, any two orbifold group actions that differ by conjugation
by a constant map can be related by gauge transformation.
Thus, on the canonical trivial bundle with trivial connection,
distinct orbifold group actions are defined by
elements of $\mbox{Hom}(\Gamma, G)/G$, where modding out $G$ is
done by conjugation.

Thus, on canonically trivial bundles with zero connection,
we find a family of orbifold group actions defined by 
$\mbox{Hom}(\Gamma, G)/G$.  This result is often used in
discussions of heterotic orbifolds -- for example, it can be
found in\footnote{In that reference, the group
$\mbox{Hom}(\Gamma, G)/G$ is described in a rather obscure
fashion.  Specifically, it is described in terms of root and weight
lattices, 
and only for the special case $\Gamma = {\bf Z}_n$.}
\cite{dixonthesis}.

We should emphasize that the occurence of $\mbox{Hom}(\Gamma, G)/G$ above
for nonabelian $G$ is much more restrictive than its occurrence
for abelian $G$.  For nonabelian $G$, we have found $\mbox{Hom}(
\Gamma, G)/G$ only for the special case of trivial principal $G$-bundles
with zero connection.  For abelian $G$, 
$\mbox{Hom}(\Gamma, G)/G = \mbox{Hom}(\Gamma, G)$ is ubiquitous -- 
for abelian $G$, elements of this group define differences between
orbifold group actions on any\footnote{Assuming, as always,
that the principal $G$-bundle with connection admits an action of the
orbifold group $\Gamma$.} principal $G$-bundle with connection.

\section{Orbifold group actions on induced gerbes}   \label{orbgrb}

Before we can understand orbifold group actions on heterotic
$B$ fields, we first need to work out the orbifold group
actions on the \v{C}ech-de Rham cocycles associated to
$\mbox{Tr } F \wedge F$ and $\mbox{Tr } R \wedge R$,
as induced by orbifold group actions on the corresponding
bundles with connection.

As mentioned previously, the \v{C}ech-de Rham cocycle description of
$\mbox{Tr } F \wedge F$ and $\mbox{Tr } R \wedge R$ is describing 
the connection on an associated 2-gerbe.  The orbifold group action
on the gauge and tangent bundles will induce an orbifold group
action on these associated 2-gerbes with connection.
Now, orbifold group actions on 2-gerbes with connection were
previously studied in \cite{cdt}, so we can borrow the results
of that paper to write, in general:
\begin{eqnarray*}
g^* \omega_3^{\alpha} & = & \omega_3^{\alpha} \: + \:
d \Lambda^{(2)}(g)^{\alpha} \\
g^* \omega_2^{\alpha \beta} & = & \omega_2^{\alpha \beta} \: + \:
d \Lambda^{(1)}(g)^{\alpha \beta} \: + \:
\Lambda^{(2)}(g)^{\alpha} \: - \: \Lambda^{(2)}(g)^{\beta} \\
g^* \omega_1^{\alpha \beta \gamma} & = & \omega_1^{\alpha \beta \gamma}
\: + \: d \log \nu^g_{\alpha \beta \gamma} \: + \:
\Lambda^{(1)}(g)^{\alpha \beta} \: + \:
\Lambda^{(1)}(g)^{\beta \gamma} \: + \:
\Lambda^{(1)}(g)^{\gamma \alpha} \\
g^* h_{\alpha \beta \gamma \delta} & = &
\left( h_{\alpha \beta \gamma \delta} \right) \,
\left( \nu^g_{\beta \gamma \delta} \right) \,
\left( \nu^g_{\alpha \gamma \delta} \right)^{-1} \,
\left( \nu^g_{\alpha \beta \delta} \right) \,
\left( \nu^g_{\alpha \beta \gamma} \right)^{-1} \\
 \, & \, & \, \\
\Lambda^{(2)}(g_1 g_2)^{\alpha} & = & \Lambda^{(2)}(g_2)^{\alpha} \: + \:
g_2^* \Lambda^{(2)}(g_1)^{\alpha} \: + \:
d \Lambda^{(3)}(g_1, g_2)^{\alpha} \\
\Lambda^{(1)}(g_1 g_2)^{\alpha \beta} & = &
\Lambda^{(1)}(g_2)^{\alpha \beta} \: + \:
g_2^* \Lambda^{(1)}(g_1)^{\alpha \beta} \: - \:
\Lambda^{(3)}(g_1, g_2)^{\alpha} \: + \:
\Lambda^{(3)}(g_1, g_2)^{\beta}  \\
 & & \: \: \: - \:
d \log \lambda^{g_1, g_2}_{\alpha \beta} \\
\Lambda^{(3)}(g_2, g_3)^{\alpha} \: + \:
\Lambda^{(3)}(g_1, g_2 g_3)^{\alpha} & = &
g_3^* \Lambda^{(3)}(g_1, g_2)^{\alpha} \: + \:
\Lambda^{(3)}(g_1 g_2, g_3)^{\alpha} \: + \:
d \log \gamma^{g_1, g_2, g_3}_{\alpha} \\
\, & \, & \, \\
\nu^{g_1 g_2}_{\alpha \beta \gamma} & = &
\left( \nu^{g_2}_{\alpha \beta \gamma} \right) \,
\left( g_2^* \nu^{g_1}_{\alpha \beta \gamma} \right) \,
\left( \lambda^{g_1, g_2}_{\alpha \beta} \right) \,
\left( \lambda^{g_1, g_2}_{\beta \gamma} \right) \,
\left( \lambda^{g_1, g_2}_{\gamma \alpha} \right) \\
\left( \lambda^{g_1 g_2, g_3}_{\alpha \beta} \right) \,
\left( g_3^* \lambda^{g_1, g_2}_{\alpha \beta} \right) & = &
\left( \lambda^{g_1, g_2 g_3}_{\alpha \beta} \right) \,
\left( \lambda^{g_2, g_3}_{\alpha \beta} \right) \,
\left( \gamma^{g_1, g_2, g_3}_{\alpha} \right) \,
\left( \gamma^{g_1, g_2, g_3}_{\beta} \right)^{-1} \\
\left( \gamma^{g_1, g_2, g_3 g_4}_{\alpha} \right) \,
\left( \gamma^{g_1 g_2, g_3, g_4}_{\alpha} \right) & = &
\left( \gamma^{g_1, g_2 g_3, g_4}_{\alpha} \right) \,
\left( \gamma^{g_2, g_3, g_4}_{\alpha} \right) \,
\left( g_4^* \gamma^{g_1, g_2, g_3}_{\alpha} \right)
\end{eqnarray*}
for some forms $\Lambda^{(2)}(g)^{\alpha}$,
$\Lambda^{(1)}(g)^{\alpha \beta}$, $\Lambda^{(3)}(g_1, g_2)^{\alpha}$,
$\nu^g_{\alpha \beta \gamma}$, $\lambda^{g_1, g_2}_{\alpha \beta}$,
and $\gamma^{g_1, g_2, g_3}_{\alpha}$ which define the orbifold
group action on the corresponding principal bundles with connection.

As a much simpler example, it is very straightforward to work
out the orbifold group action induced on the 0-gerbe (determinant
bundle) associated to some principal bundle with connection.
Recall that the 0-gerbe with connection has curvature $\mbox{Tr } F$,
local connections $\mbox{Tr } A^{\alpha}$, and transition functions
$\det g_{\alpha \beta}$.
Also recall that the orbifold group action on a principal $G$-bundle
with connection is described by functions $\gamma^g_{\alpha}$,
where
\begin{eqnarray*}
g^* A^{\alpha} & = & \left( \gamma^g_{\alpha} \right) 
\, A^{\alpha} \, \left( \gamma^g_{\alpha} \right)^{-1}
\: + \: \left( \gamma^g_{\alpha} \right)
d \left( \gamma^g_{\alpha} \right)^{-1} \\
g^* g_{\alpha \beta} & = & \left( \gamma^g_{\alpha} \right) \,
\left( g_{\alpha \beta} \right) \,
\left( \gamma^g_{\beta} \right)^{-1} \\
\gamma^{g_1 g_2}_{\alpha} & = &
\left( g_2^* \gamma^{g_1}_{\alpha} \right) \,
\left( \gamma^{g_2}_{\alpha} \right)
\end{eqnarray*}
From this description, it is easy to compute that
\begin{eqnarray*}
g^* \mbox{Tr } F & = & \mbox{Tr }F \\
g^* \mbox{Tr } A^{\alpha} & = &
\mbox{Tr }A^{\alpha} \: + \: \mbox{Tr } \left( 
\left( \gamma^g_{\alpha} \right) \, d \left( \gamma^g_{\alpha} \right)^{-1}
\right) \\
 & = & \mbox{Tr } A^{\alpha} \: + \: d \log \left( \det \gamma^g_{\alpha} 
\right) \\
\det \gamma^{g_1 g_2}_{\alpha} & = & 
\left( \det \gamma^{g_2}_{\alpha} \right) \,
\left( g_2^* \det \gamma^{g_1}_{\alpha} \right)
\end{eqnarray*}
so we see explicitly that the orbifold group action on a principal
$G$-bundle with connection defines an orbifold group action on the
associated 0-gerbe (determinant bundle) with connection.

\section{Orbifold group actions on heterotic $B$-fields}   \label{orbhetB}

Now that we have described heterotic $B$ fields on local coordinate
patches, and described the orbifold group actions induced on Chern-Simons
forms by orbifold group actions on principal bundles with connection,
we are finally ready to work out orbifold group actions on 
heterotic $B$ fields.

First, recall that in the Green-Schwarz mechanism, gauge transformations
of the bundle which induce
\begin{displaymath}
\omega_3^{\alpha} \: \mapsto \: \omega_3^{\alpha} \: - \:
\mbox{Tr } \left( d \Lambda \wedge d A^{\alpha} \right)
\end{displaymath}
the $B$ field transforms as
\begin{displaymath}
B^{\alpha} \: \mapsto \: B^{\alpha} \: + \:
\mbox{Tr } \left( \Lambda d A^{\alpha} \right)
\end{displaymath}
(so that $H$ remains invariant).
From this fact and the fact that under the action of the orbifold
group,
\begin{displaymath}
g^* \omega_3^{\alpha} \: = \:
\omega_3^{\alpha} \: + \: d \Lambda^{(2)}(g)^{\alpha}
\end{displaymath}
we see that, in general, 
\begin{equation}
g^* B^{\alpha} \: = \: B^{\alpha} \: - \:
\Lambda^{(2,F)}(g)^{\alpha} \: + \:
\Lambda^{(2,R)}(g)^{\alpha} \: + \:
d \Lambda^{(1,B)}(g)^{\alpha}
\end{equation}
for some local one-forms $\Lambda^{(1,B)}(g)^{\alpha}$.

Also note that this implies that $g^* H = H$.
In fact, we should have expected this -- since $H$ has no gauge 
transformations, any well-defined orbifold group action must map
$H$ back into precisely itself.

From the fact that
\begin{displaymath}
B^{\alpha} \: - \: B^{\beta} \: = \:
d A^{\alpha \beta} \: - \: \omega_{2, F}^{\alpha \beta} \: + \:
\omega_{2, R}^{\alpha \beta}
\end{displaymath}
we can derive that 
\begin{equation}
g^* A^{\alpha \beta} \: = \:
A^{\alpha \beta} \: + \:
\Lambda^{(1,B)}(g)^{\alpha} \: - \:
\Lambda^{(1,B)}(g)^{\beta} \: + \:
\Lambda^{(1,F)}(g)^{\alpha \beta} \: - \:
\Lambda^{(1,R)}(g)^{\alpha \beta} \: + \:
d \log \kappa^g_{\alpha \beta}
\end{equation}
for some local function $\kappa^g_{\alpha \beta}$.

From the fact that
\begin{displaymath}
A^{\alpha \beta} \: + \: A^{\beta \gamma} \: + \: A^{\gamma \alpha}
\: = \:
\omega_{1, F}^{\alpha \beta \gamma} \: - \:
\omega_{1, R}^{\alpha \beta \gamma} \: + \:
d \log h^B_{\alpha \beta \gamma}
\end{displaymath}
we can derive that
\begin{equation}
g^* h^B_{\alpha \beta \gamma} \: = \:
\left( h^B_{\alpha \beta \gamma} \right) \,
\left( \nu^{F g}_{\alpha \beta \gamma} \right)^{-1} \,
\left( \nu^{R g}_{\alpha \beta \gamma} \right) \,
\left( \kappa^g_{\alpha \beta} \right) \,
\left( \kappa^g_{\beta \gamma} \right) \,
\left( \kappa^g_{\gamma \alpha} \right)
\end{equation}

From expanding $(g_1 g_2)^* h^B_{\alpha \beta \gamma}$ in two different
ways, we find that
\begin{equation}
\left( \lambda^{F g_1, g_2}_{\alpha \beta} \right)^{-1} \,
\left( \lambda^{R g_1, g_2}_{\alpha \beta} \right) \,
\left( \kappa^{g_1 g_2}_{\alpha \beta} \right) \: = \:
\left( \kappa^{g_2}_{\alpha \beta} \right) \,
\left( g_2^* \kappa^{g_1}_{\alpha \beta} \right) \,
\left( h^{g_1, g_2}_{\alpha} \right) \,
\left( h^{g_1, g_2}_{\beta} \right)^{-1} 
\end{equation}
for some local functions $h^{g_1, g_2}_{\alpha}$.

From writing $\kappa^{g_1 g_2 g_3}_{\alpha \beta}$ in two different
ways, we find that
\begin{equation}
\left( \gamma^{F g_1, g_2, g_3}_{\alpha} \right) \,
\left( \gamma^{R g_1, g_2, g_3}_{\alpha} \right)^{-1} \,
\left( h^{g_1 g_2, g_3}_{\alpha} \right) \,
\left( g_3^* h^{g_1, g_2}_{\alpha} \right) \: = \:
\left( h^{g_1, g_2 g_3}_{\alpha} \right) \,
\left( h^{g_2, g_3}_{\alpha} \right)
\end{equation}

From expanding $(g_1 g_2)^* B^{\alpha}$ in two different ways, we find
\begin{equation}
- d \Lambda^{(3, F)}(g_1, g_2)^{\alpha} \: + \:
d \Lambda^{(3, R)}(g_1, g_2)^{\alpha} \: + \:
d \Lambda^{(1, B)}(g_1 g_2)^{\alpha} \: = \:
d \Lambda^{(1, B)}(g_2)^{\alpha} \: + \:
g_2^* d \Lambda^{(1, B)}(g_1)^{\alpha}
\end{equation}
and from expanding $(g_1 g_2)^* A^{\alpha \beta}$ in two different ways,
we find
\begin{eqnarray}
\delta \left[ \Lambda^{(1,B)}(g_1 g_2)^{\alpha} \: - \:
\Lambda^{(3, F)}(g_1, g_2)^{\alpha} \right. & + & \left.
\Lambda^{(3, R)}(g_1, g_2)^{\alpha} \: + \:
d \log h^{g_1, g_2}_{\alpha} \right] \nonumber \\
& = &
\delta \left[ \Lambda^{(1,B)}(g_2)^{\alpha} \: + \:
g_2^* \Lambda^{(1,B)}(g_1)^{\alpha} \right]
\end{eqnarray}
which we combine to conclude that
\begin{equation}
\Lambda^{(1, B)}(g_1 g_2)^{\alpha} \: - \:
\Lambda^{(3, F)}(g_1, g_2)^{\alpha} \: + \:
\Lambda^{(3, R)}(g_1, g_2)^{\alpha} \: + \:
d \log h^{g_1, g_2}_{\alpha} \: = \:
\Lambda^{(1,B)}(g_2)^{\alpha} \: + \:
g_2^* \Lambda^{(1, B)}(g_1)^{\alpha}
\end{equation}

To summarize, we have discovered that an orbifold group
action on a heterotic $B$ field is defined by
\begin{eqnarray*}
g^* H & = & H \\
g^* B^{\alpha} & = & B^{\alpha} \: - \:
\Lambda^{(2,F)}(g)^{\alpha} \: + \:
\Lambda^{(2,R)}(g)^{\alpha} \: + \:
d \Lambda^{(1,B)}(g)^{\alpha} \\
g^* A^{\alpha \beta} & = &
A^{\alpha \beta} \: + \:
\Lambda^{(1,B)}(g)^{\alpha} \: - \:
\Lambda^{(1,B)}(g)^{\beta}  \\
& & \: + \:
\Lambda^{(1,F)}(g)^{\alpha \beta} \: - \:
\Lambda^{(1,R)}(g)^{\alpha \beta}  
 \: + \:
d \log \kappa^g_{\alpha \beta} \\
g^* h^B_{\alpha \beta \gamma} & = &
\left( h^B_{\alpha \beta \gamma} \right) \,
\left( \nu^{F g}_{\alpha \beta \gamma} \right)^{-1} \,
\left( \nu^{R g}_{\alpha \beta \gamma} \right) \,
\left( \kappa^g_{\alpha \beta} \right) \,
\left( \kappa^g_{\beta \gamma} \right) \,
\left( \kappa^g_{\gamma \alpha} \right) \\
\, & \, & \, \\
\left( \lambda^{F g_1, g_2}_{\alpha \beta} \right)^{-1} \,
\left( \lambda^{R g_1, g_2}_{\alpha \beta} \right) \,
\left( \kappa^{g_1 g_2}_{\alpha \beta} \right) & = &
\left( \kappa^{g_2}_{\alpha \beta} \right) \,
\left( g_2^* \kappa^{g_1}_{\alpha \beta} \right) \,
\left( h^{g_1, g_2}_{\alpha} \right) \,
\left( h^{g_1, g_2}_{\beta} \right)^{-1} \\
\left( h^{g_1 g_2, g_3}_{\alpha} \right) \,
\left( g_3^* h^{g_1, g_2}_{\alpha} \right) & = &
\left( h^{g_1, g_2 g_3}_{\alpha} \right) \,
\left( h^{g_2, g_3}_{\alpha} \right) \,
\left( \gamma^{F g_1, g_2, g_3}_{\alpha} \right)^{-1} \,
\left( \gamma^{R g_1, g_2, g_3}_{\alpha} \right) \\
\Lambda^{(1, B)}(g_1 g_2)^{\alpha} 
\: + \:
d \log h^{g_1, g_2}_{\alpha} & = &
\Lambda^{(3, F)}(g_1, g_2)^{\alpha} \: - \:
\Lambda^{(3, R)}(g_1, g_2)^{\alpha} \: + \:
\Lambda^{(1,B)}(g_2)^{\alpha} \\
& & \: + \:
g_2^* \Lambda^{(1, B)}(g_1)^{\alpha}
\end{eqnarray*}
for some $\Lambda^{(1,B)}(g)^{\alpha}$, $\kappa^g_{\alpha \beta}$,
and $h^{g_1, g_2}_{\alpha}$ introduced to define the orbifold group action
on the heterotic $B$ field.
Note that this is the same set of data needed to define an orbifold
group action on a $B$ field for the case $d H = 0$ \cite{dt1,dt2,dt3};
the difference in the present case is that the orbifold group
action is warped by the interaction with the gauge and tangent bundles.

\section{Differences between orbifold group actions}   \label{diffs}

In \cite{dt1,dt2,dt3}, the group $H^2(\Gamma, U(1))$ was recovered
when describing the differences between orbifold group actions
on $B$ fields such that $d H = 0$.
With that in mind, we shall now examine the differences between
orbifold group actions on heterotic $B$ fields.

Assume the orbifold group actions on the gauge and tangent bundles
are fixed.  Let the data defining the two orbifold group
actions on the heterotic $B$ field be distinguished by an overline.
Define
\begin{eqnarray*}
T^g_{\alpha \beta} & = & \frac{ \kappa^g_{\alpha \beta} }{
\overline{\kappa}^g_{\alpha \beta} } \\
A(g)^{\alpha} & = & \overline{\Lambda}^{(1, B)}(g)^{\alpha} \: - \:
\Lambda^{(1, B)}(g)^{\alpha} \\
\omega^{g_1, g_2}_{\alpha} & = & \frac{ h^{g_1, g_2}_{\alpha} }{
\overline{h}^{g_1, g_2}_{\alpha} } 
\end{eqnarray*}

From the expressions
\begin{eqnarray*}
g^* B^{\alpha} & = & B^{\alpha} \: - \: \Lambda^{(2, F)}(g)^{\alpha} \: + \:
\Lambda^{(2, R)}(g)^{\alpha} \: + \: d \Lambda^{(1, B)}(g)^{\alpha} \\
 & = & B^{\alpha} \: - \: \Lambda^{(2, F)}(g)^{\alpha} \: + \:
\Lambda^{(2, R)}(g)^{\alpha} \: + \: d \overline{ \Lambda }^{(1, B)}(g)^{
\alpha} 
\end{eqnarray*}
we see that
\begin{equation}
d A(g)^{\alpha} \: = \: 0
\end{equation}

From writing $g^* A^{\alpha \beta}$ in two different ways, we find that
\begin{equation}
A(g)^{\alpha} \: - \: A(g)^{\beta} \: = \: d \log T^g_{\alpha \beta}
\end{equation}

From writing $g^* h^B_{\alpha \beta \gamma}$ in two different ways,
we find that
\begin{equation}
\left( T^g_{\alpha \beta} \right) \,
\left( T^g_{\beta \gamma} \right) \,
\left( T^g_{\gamma \alpha} \right) 
\: = \: 1
\end{equation}

From the equations above, we see that the $T^g_{\alpha \beta}$ are
transition functions for a principal $U(1)$ bundle with connection
defined by $A(g)^{\alpha}$, and that that connection is flat.

By dividing the expressions for $\kappa^{g_1 g_2}_{\alpha \beta}$
and $\overline{\kappa}^{g_1, g_2}_{\alpha \beta}$, we find that
\begin{equation}
T^{g_1 g_2}_{\alpha \beta} \: = \:
\left( T^{g_2}_{\alpha \beta} \right) \,
\left( g_2^* T^{g_1}_{\alpha \beta} \right) \,
\left( \omega^{g_1, g_2}_{\alpha} \right) \,
\left( \omega^{g_1, g_2}_{\beta} \right)^{-1}
\end{equation}

From subtracting the expressions for $\Lambda^{(1, B)}(g)^{\alpha}$
and $\overline{\Lambda}^{(1, B)}(g)^{\alpha}$, we find that
\begin{equation}
A(g_1 g_2)^{\alpha} \: - \: d \log \omega^{g_1, g_2}_{\alpha}
\: = \:
A(g_2)^{\alpha} \: + \: g_2^* A(g_1)^{\alpha}
\end{equation}

These two expressions tell us that the $\omega^{g_1, g_2}_{\alpha}$
define connection-preserving bundle isomorphisms
\begin{displaymath}
\omega^{g_1, g_2}: \: T^{g_2} \otimes g_2^* T^{g_1} \: 
\longrightarrow \:
T^{g_1 g_2}
\end{displaymath}

Finally, by dividing the expressions
\begin{eqnarray*}
\left( \gamma^{F g_1, g_2, g_3}_{\alpha} \right) \,
\left( \gamma^{R g_1, g_2, g_3}_{\alpha} \right)^{-1} \,
\left( h^{g_1 g_2, g_3}_{\alpha} \right) \,
\left( g_3^* h^{g_1, g_2}_{\alpha} \right) & = &
\left( h^{g_1, g_2 g_3}_{\alpha} \right) \,
\left( h^{g_2, g_3}_{\alpha} \right) \\
\left( \gamma^{F g_1, g_2, g_3}_{\alpha} \right) \,
\left( \gamma^{R g_1, g_2, g_3}_{\alpha} \right)^{-1} \,
\left( \overline{h}^{g_1 g_2, g_3}_{\alpha} \right) \,
\left( g_3^* \overline{h}^{g_1, g_2}_{\alpha} \right) & = &
\left( \overline{h}^{g_1, g_2 g_3}_{\alpha} \right) \,
\left( \overline{h}^{g_2, g_3}_{\alpha} \right) 
\end{eqnarray*}
we find that
\begin{equation}
\left( \omega^{g_1 g_2, g_3}_{\alpha} \right) \,
\left( g_3^* \omega^{g_1, g_2}_{\alpha} \right) \: = \:
\left( \omega^{g_1, g_2 g_3}_{\alpha} \right) \,
\left( \omega^{g_2, g_3}_{\alpha} \right)
\end{equation}

This means that the connection-preserving bundle morphisms
$\omega^{g_1, g_2}$ must make the following diagram commute:
\begin{equation}    \label{omegacocycle}
\begin{array}{ccc}
T^{g_3} \otimes g_3^* \left( \, T^{g_2} \otimes g_2^* T^{g_1} \right)
& \stackrel{ \omega^{g_1, g_2} }{ \longrightarrow } &
T^{g_3} \otimes g_3^* T^{g_1 g_2} \\
\makebox[0pt][r]{ $\scriptstyle{  \omega^{g_2, g_3} }$} \downarrow
& & \downarrow \makebox[0pt][l]{
$\scriptstyle{ \omega^{g_1 g_2, g_3} }$ } \\
T^{g_2 g_3} \otimes (g_2 g_3)^* T^{g_1} &
\stackrel{ \omega^{g_1, g_2 g_3} }{\longrightarrow } &
T^{g_1 g_2 g_3}
\end{array}
\end{equation}

So far we have recovered the fact that the difference between
two orbifold group actions on heterotic $B$ fields
(with fixed orbifold group actions on the gauge and tangent bundles)
is defined by the same data as for $B$ fields such that $d H = 0$
\cite{dt1,dt2,dt3}:  namely, pairs $(T^g, \omega^{g_1, g_2})$
of bundles $T^g$ with flat connection and connecting morphisms 
$\omega^{g_1, g_2}$ making
diagram~(\ref{omegacocycle}) commute.

Also, orbifold group actions on $B$ fields are subject to the
same equivalences as in \cite{dt1,dt2,dt3}.
If $\kappa_g: T^g \rightarrow T'^g$ is a connection-preserving
isomorphism of principal $G$-bundles, then we can replace the
data $(T^g, \omega^{g_1, g_2})$ with the data
$\left( T'^g, \kappa_{g_1 g_2} \circ \omega^{g_1, g_2} \circ
\left( \kappa_{g_2} \otimes g_2^* \kappa_{g_1} \right)^{-1} \right)$.

Since the differences between orbifold group actions
on heterotic $B$ fields are defined by precisely the same
data as for type II $B$ fields \cite{dt1,dt2,dt3}, we recover
the group
$H^2(\Gamma, U(1))$ as well as the twisted-sector phases of
\cite{vafa1} in precisely the same fashion as \cite{dt1,dt2,dt3}.

\section{Conclusions}

In this paper we have outlined a purely mathematical
understanding of discrete torsion for heterotic $B$ fields,
as opposed to type II $B$ fields,
thereby filling a gap present in the earlier work \cite{dt1,dt2,dt3}. 
Specifically, after working out a gerbe-like description of
heterotic $B$ fields, and after discussing orbifold group
actions on principal $G$-bundles with connection for non-abelian $G$,
we use a self-consistent bootstrap (in the style of \cite{dt3}) to
construct orbifold group actions on $B$ fields.  Discrete torsion
arises in the same fashion as in \cite{dt1,dt2,dt3}, namely in terms
of the difference between orbifold group actions.

As in \cite{dt1,dt2,dt3}, the results in this paper do not assume
that the orbifold group acts freely.  Also as in \cite{dt1,dt2,dt3},
we do not assume the heterotic $B$ field has vanishing curvature
(though, as in \cite{dt1,dt2,dt3}, one needs to check that orbifold
group actions on a given field configuration actually exist before
attempting to formally classify them).

Finally, as in \cite{dt1,dt2,dt3}, our analysis does not assume
any features of string theory.  As in \cite{dt1,dt2,dt3},
discrete torsion can be understood in a purely mathematical
framework, without any reference to string theory.  In other words,
there is nothing ``inherently stringy'' about discrete torsion.

One loose end we have had difficulty tying up involves the
level-matching conditions of heterotic orbifolds.
We strongly suspect that satisfying the level-matching conditions
is equivalent to the statement that the orbifold group actions
on the gauge and tangent bundles are consistent with the orbifold
group action on the heterotic $B$ field.  In other words,
we suspect the level-matching condition is equivalent to demanding
that the orbifold group action on the heterotic $B$ field be well-defined.
Unfortunately, we have not yet been able to show this rigorously.

\section{Acknowledgements}

We would like to thank  P.~Aspinwall, A.~Knutson, D.~Morrison,
and R.~Plesser for useful conversations.

This research was partially supported by National Science Foundation
grant number DMS-0074072.

\end{document}